%% file: main.tex
\documentclass[10pt,conference]{IEEEtran}

\usepackage{cite}
\usepackage{amsmath,amssymb,amsfonts}
\usepackage{algorithmic}
\usepackage{graphicx}
\usepackage{textcomp}
\usepackage{url}
\usepackage{xcolor}
\usepackage{balance}
\usepackage[hidelinks]{hyperref}
\usepackage{eso-pic}

\expandafter\ifx\csname iflatexml\endcsname\relax
  \expandafter\newif\csname iflatexml\endcsname
\fi

\input{body/macro.tex}

\def\BibTeX{{\rm B\kern-.05em{\sc i\kern-.025em b}\kern-.08em
    T\kern-.1667em\lower.7ex\hbox{E}\kern-.125emX}}

\begin{document}
\AddToShipoutPictureFG*{%
    \AtTextLowerLeft{%
        \put(0,\LenToUnit{-1.0cm}){%
            \parbox[t]{\textwidth}{\scriptsize
                \copyright~2026 IEEE. Personal use of this material is permitted.
                Permission from IEEE must be obtained for all other uses, in any
                current or future media, including reprinting/republishing this
                material for advertising or promotional purposes, creating new
                collective works, for resale or redistribution to servers or lists,
                or reuse of any copyrighted component of this work in other works.}}}}

\title{Symmetry-Oriented Testing Framework for Variational Quantum Algorithms}


\author{
    \IEEEauthorblockN{Masaomi Yamaguchi}
    \IEEEauthorblockA{\textit{Fujitsu Limited} \\
        Kawasaki, Kanagawa, Japan \\
        \textit{The University of Electro-Communications} \\
        Chofu, Tokyo, Japan \\
        y.masaomi@fujitsu.com}
    \and
    \IEEEauthorblockN{Fuyuki Ishikawa}
    \IEEEauthorblockA{\textit{National Institute of Informatics} \\
        Chiyoda-ku, Tokyo, Japan \\
        f-ishikawa@nii.ac.jp}
}

\maketitle

\iflatexml
    \noindent\copyright~2026 IEEE. Personal use of this material is permitted.
    Permission from IEEE must be obtained for all other uses, in any current or
    future media, including reprinting/republishing this material for advertising
    or promotional purposes, creating new collective works, for resale or
    redistribution to servers or lists, or reuse of any copyrighted component of
    this work in other works.
\fi

\input{body/abstract.tex}
\input{body/introduction.tex}
\input{body/background.tex}
\input{body/related_work.tex}
\input{body/motivating_example.tex}
\input{body/proposed_method.tex}
\input{body/evaluation.tex}
\input{body/discussion.tex}
\input{body/conclusion.tex}
\input{body/data_availability.tex}
\input{body/acknowledgment.tex}
\balance
\bibliographystyle{IEEEtran}
\bibliography{references}

\end{document}

%% file: body/macro.tex


\newenvironment{tarray}[2][c]{%
\settowidth{\dimen1}{${} = {}$}%
\setlength{\arraycolsep}{0.125\dimen1}
\hspace{-0.125\dimen1}\array[#1]{#2}\relax
}
{%
\endarray\hspace{-0.125\dimen1}
}

\newcommand{\gate}[1]{\ensuremath{\mathit{#1}}}

\newcommand{\bra}[1]{\left\langle{#1}\right\rvert}
\newcommand{\ket}[1]{\left\lvert{#1}\right\rangle}

\newif\ifdraft
\draftfalse

\ifdraft
  \definecolor{darkpastelgreen}{rgb}{0.01, 0.75, 0.24}
  \newcommand{\revised}[1] {%
    \textcolor{darkpastelgreen}{#1}%
  }
  
  \newenvironment{revisedblock}{%
    \color{darkpastelgreen}%
  }{%
    \ignorespacesafterend%
  }
\else
  \definecolor{darkpastelgreen}{rgb}{0.01, 0.75, 0.24}
  \newcommand{\revised}[1] {%
    \textcolor{black}{#1}%
  }
  
  \newenvironment{revisedblock}{%
    \color{black}%
  }{%
    \ignorespacesafterend%
  }
\fi

%% file: body/abstract.tex
\begin{abstract}
  Testing quantum software is challenging because quantum states cannot be directly inspected. Variational quantum algorithms (VQAs) are a particularly important target as they are among the most promising approaches for near-term noisy quantum hardware. Many testing methods rely on classically computed results, which conflicts with the regime where quantum advantage is expected. Although prior studies mitigate this oracle problem, the core difficulty remains that existing approaches do not verify whether the execution deviates from the intended problem or whether the obtained solution is physically plausible.

  We propose a symmetry-oriented testing framework for VQAs that addresses this gap by using operators commuting with the target Hamiltonian as test specifications. The framework provides two complementary diagnostics: a commutator-based health check that monitors whether each execution step during optimization remains consistent with the symmetry of the target Hamiltonian, and a variance-based acceptance check that assesses the converged solution by examining whether the state is an eigenstate of the symmetry operators. Both diagnostics derive from the problem structure and do not require the exact solution.

  We evaluate the framework on representative user- and system-side faults, across multiple target systems, and under injected noise. Results show that the two diagnostics detect faulty behavior, provide informative signals of solution quality, and remain useful under noisy execution.
\end{abstract}


%% file: body/introduction.tex
\section{Introduction}
\label{sec:introduction}

Quantum computing is expected to support important applications such as database search, optimization, chemistry simulation, and machine learning \cite{Grover1996,Shor1997PrimeFactorization,AspuruGuzik2005PEA,Peruzzo2014VQE}. As these applications move toward regimes where quantum computers are expected to outperform classical computers, an increasingly important question is how to validate the plausibility of the obtained solution. Among the available validation techniques, testing remains attractive because of its comparatively low cost and its compatibility with practical development workflows. However, testing quantum software is inherently difficult because the expected behavior of a program is encoded in a quantum state that cannot be directly inspected.

Variational quantum algorithms (VQAs) are a particularly important target for testing methods because they are among the most promising approaches for current noisy quantum hardware and may also serve as a route toward \revised{practical quantum advantage} \cite{Peruzzo2014VQE}. At the same time, a VQA can produce an incorrect result for many reasons, including noise, compiler defects, user-side implementation mistakes, insufficient ansatz expressiveness, and convergence to poor local optima.

The central challenge is the test oracle \cite{Baar2015ClassicalTestOracle}. Many existing approaches assume that the correct output is already available, for example through an analytical solution or a classical simulator. This assumption breaks down in the regime where quantum advantage is expected, because the correct solution cannot be obtained classically. Alternative approaches reduce this dependence by, for example, checking selected state properties through assertions \cite{Huang2019StatisticalAssertions,Li2020Projection}, generating tests based on coverage criteria \cite{Wang2022Quito} or verifying user-specified invariants~\cite{honarvar2020PropertyTest,Muqeet2024QOPS}, or comparing related executions to detect platform-level defects \cite{Wang2022QDiff,Paltenghi2023MorphQ}. These methods are effective for their respective purposes, but assessing the behavior at each execution step during optimization and the converged solution still requires detailed knowledge of the expected states at the relevant program points.

To address this challenge, we propose a symmetry-oriented testing framework (abbreviated as SOT Framework) for VQAs. The key idea is to use symmetries of the target Hamiltonian as test specifications, represented by operators that commute with the target Hamiltonian. The framework provides two complementary diagnostics: a commutator-based health check that monitors whether each execution step during optimization remains consistent with the symmetry of the target Hamiltonian, and a variance-based acceptance check that assesses the converged solution by examining whether the state is an eigenstate of the symmetry operators. Both diagnostics derive from the problem structure and do not require the exact solution.

To evaluate the proposed framework, we study the following research questions:
\begin{itemize}
  \item \textbf{RQ1:} Can the proposed framework provide useful fault-detection signals for concrete fault scenarios in VQA workflows without relying on the exact solution as a test oracle?
  \item \textbf{RQ2:} Does the proposed acceptance check show a meaningful correlation with solution quality across diverse target systems?
  \item \textbf{RQ3:} Can the proposed acceptance check serve as a practical filter for low-quality VQA outcomes?
  \item \textbf{RQ4:} Can the proposed framework provide useful diagnostic signals under noisy execution conditions?
\end{itemize}

The main contributions of this paper are as follows:
\begin{itemize}
  \item We propose a symmetry-oriented testing framework for VQAs that uses symmetry of the target Hamiltonian as test specifications, enabling oracle-free testing of VQA workflows.

  \item We instantiate the framework as two complementary diagnostics: a commutator-based health check for each execution step during optimization and a variance-based acceptance check for the converged solution.

  \item We provide a comprehensive empirical evaluation of the proposed framework across multiple fault scenarios, correlation analyses between the acceptance check and solution quality, diverse target systems, and noisy execution conditions.
\end{itemize}

%% file: body/background.tex
\section{Background}
\label{sec:background}

\subsection{Basics of quantum computing}
\label{sec:basics}

The basic unit of quantum computation is the qubit, whose pure state is a unit vector up to a global phase in $\mathbb{C}^2$. With computational-basis states $\ket{0} = (1,0)^\top$ and $\ket{1} = (0,1)^\top$, any pure state can be written as $\ket{\psi} = \alpha \ket{0} + \beta \ket{1}$ with $\alpha, \beta \in \mathbb{C}$ and $|\alpha|^2 + |\beta|^2 = 1$. We write $\bra{\psi}$ for the Hermitian conjugate of $\ket{\psi}$. The state of $n$ qubits is a unit vector up to a global phase in $(\mathbb{C}^{2})^{\otimes n}$, and $\otimes$ is omitted when clear.

\begin{figure}[t]
  \centering
  \includegraphics[scale=0.4]{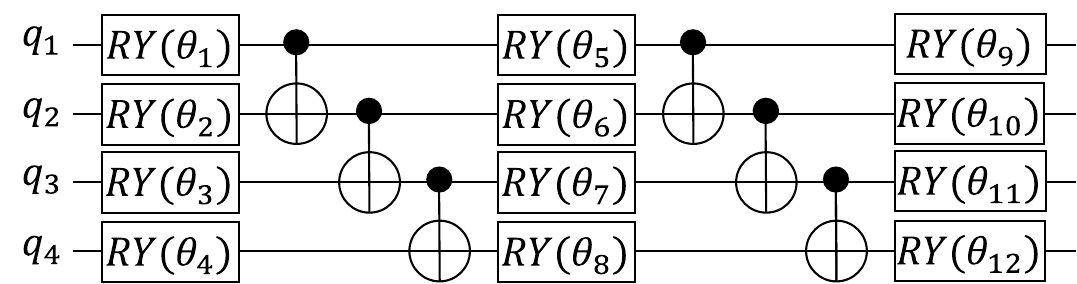}
  \caption{An example circuit consisting of $\gate{R}_y$ and $\gate{CX}$ layers.}
  \label{fig:background-circuit}
\end{figure}

A \emph{quantum circuit} is a sequence of unitary operators (\emph{quantum gates}) acting on quantum states; applying a gate $U$ to $\ket{\psi}$ yields $U\ket{\psi}$. The single-qubit Pauli operators are
\begin{equation}
  X = \begin{pmatrix}0 & 1 \\ 1 & 0\end{pmatrix},
  \quad
  Y = \begin{pmatrix}0 & -i \\ i & 0\end{pmatrix},
  \quad
  Z = \begin{pmatrix}1 & 0 \\ 0 & -1\end{pmatrix}
\end{equation}
Fig.~\ref{fig:background-circuit} shows an example circuit. Here, $\gate{R}_y(\theta) = e^{-i\theta Y/2}$ and $\gate{CX} = \ket{0}\bra{0}\otimes I + \ket{1}\bra{1}\otimes X$, where $I$ is the identity operator. We also use $\gate{CZ} = \ket{0}\bra{0}\otimes I + \ket{1}\bra{1}\otimes Z$, $\gate{R}_Z(\theta) = e^{-i\theta Z/2}$, $\gate{S} = \frac{1+i}{2}I + \frac{1-i}{2}Z$, $\gate{SX} = \frac{1+i}{2}I + \frac{1-i}{2}X$, and $\gate{H} = \frac{X+Z}{\sqrt{2}}$.

The state produced by a circuit is not observed directly. Instead, information is obtained by \emph{projective measurement} of a Hermitian observable $M$. Writing its spectral decomposition as $M = \sum_i \lambda_i \Pi_i$, outcome $\lambda_i$ occurs with probability $p_i = \bra{\psi}\Pi_i\ket{\psi}$, and the post-measurement state is $\Pi_i\ket{\psi}/\sqrt{p_i}$. Because measurement is probabilistic, the circuit is executed repeatedly to estimate the \emph{expectation value}:
\begin{equation}
  \langle M \rangle_{\psi} = \sum_i p_i \lambda_i = \sum_i \lambda_i \bra{\psi}\Pi_i\ket{\psi} = \bra{\psi}M\ket{\psi}
\end{equation}

As is common in quantum computing, we assume that only $Z$ can be measured directly on each qubit. A general observable $M$ is estimated through its Pauli decomposition $M = \sum_i c_i P_i$, where each $P_i$ is a \emph{Pauli string}: a tensor product of $I$, $X$, $Y$, and $Z$. By linearity, it suffices to estimate each $\bra{\psi}P_i\ket{\psi}$ separately. For $P = \sigma_1 \otimes \cdots \otimes \sigma_n$, we apply $\gate{H}$ to qubit $j$ when $\sigma_j = X$ and $\gate{H}\gate{S}^\dagger$ when $\sigma_j = Y$, so that the product of the resulting $Z$-measurement outcomes over the non-identity factors has expectation value $\bra{\psi}P\ket{\psi}$.

For an observable $M$, the variance on $\ket{\psi}$ is
\begin{equation}
  \begin{aligned}
    \mathrm{Var}_{\psi}(M)
     & = \sum_i p_i(\lambda_i - \langle M \rangle_{\psi})^2    \\
     & = \bra{\psi} M^2 \ket{\psi} - \bra{\psi} M \ket{\psi}^2
  \end{aligned}
\end{equation}
and is estimated from the Pauli decompositions of $M$ and $M^2$.

\subsection{Variational Quantum Algorithms}
\label{sec:background-variational-quantum-algorithms}

\emph{Variational quantum algorithms} are hybrid quantum-classical procedures that optimize the parameters of a quantum circuit according to an objective function \cite{Peruzzo2014VQE,farhi2014QAOA}. In this paper, we consider the common setting in which the objective is the expectation value of a Hamiltonian $H$.

The goal is to obtain the lowest eigenvalue $E_0$ of $H$, called the ground-state energy, and its corresponding ground state. An eigenvalue is \emph{nondegenerate} when the corresponding eigenstate is unique up to a global phase. The variational principle guarantees $E_0 \le \bra{\phi} H \ket{\phi}$ for any normalized state $\ket{\phi}$, so minimizing $\bra{\phi} H \ket{\phi}$ approximates $E_0$.

Let $U(\theta)$ be a parameterized circuit with parameter vector~$\theta$, and let $\ket{\psi_0}$ be an initial state. The state $\ket{\psi(\theta)} = U(\theta)\ket{\psi_0}$ is the \emph{variational state}, and $U(\theta)$ is often called an \emph{ansatz}. Figure~\ref{fig:background-circuit} shows an example. A VQA alternates between estimating $\bra{\psi(\theta)} H \ket{\psi(\theta)}$ on a quantum device and updating~$\theta$ with a classical optimizer to approximate $E_0$.

%% file: body/related_work.tex
\section{Related Work}
\label{sec:related-work}

In quantum software testing, it is common to test programs on instances small enough for the expected output to be computed classically and used as a test oracle. This strategy becomes less practical in the regime where quantum computers are expected to provide an advantage and exact classical solutions become intractable as test oracles.

One direction is to check selected properties of the quantum state rather than the full output. Statistical assertions test whether the quantum state at a given program point satisfies a tester-specified property via statistical tests on measurement outcomes~\cite{Huang2019StatisticalAssertions}. \revised{To avoid destroying the state by measurement, subsequent work introduced non-destructive runtime assertions using} auxiliary qubits~\cite{Liu2020RuntimeAssertions,Liu2021ApproximateAssertion}. Li et al.\ formulated assertions as projection operators whose eigenstates correspond to the expected state~\cite{Li2020Projection}. Property-based testing \revised{methods \cite{honarvar2020PropertyTest, pontolillo2025qucheck} let the tester declare expected output properties and verify them} statistically over random inputs. QOPS reduces the specification burden by asking for the expected measurement-outcome distribution of one Pauli string per commuting family, deriving test oracles for all other strings in the family~\cite{Muqeet2024QOPS}. Applying any of these methods to assess VQA solution quality or to verify whether the execution remains consistent with the intended problem during optimization would require detailed knowledge of the expected states at the relevant program points. In VQA workflows, however, no such specification is available: the expected states during optimization are not known in advance, and the converged solution is precisely what the algorithm is trying to find.

A separate line of work reduces oracle dependence through automated test generation. Coverage-guided methods such as Quito~\cite{Wang2022Quito} generate test inputs that maximize structural coverage of the quantum circuit, while search-based methods such as QuSBT~\cite{wang2021TestSuiteGen,Wang2022QuSBT} use metaheuristic search to find inputs likely to expose failures. These methods are valuable for revealing faults without a complete reference output, but they target general quantum programs rather than VQA workflows specifically and do not provide a mechanism for assessing the quality of a converged VQA solution.

Another direction is to avoid specifying expected outputs altogether by comparing related executions. QDiff generates semantically equivalent circuit variants and compares their measurement-outcome distributions~\cite{Wang2022QDiff}, while MorphQ and MorphQ++ apply metamorphic transformations such as qubit reordering and null-effect insertions to check whether software stacks preserve circuit semantics~\cite{Paltenghi2023MorphQ,Linsey2024morphqplusplus}. Because these methods require no knowledge of the correct answer, they are effective for detecting platform-level defects without a classical reference. However, their diagnostic target is infrastructure correctness: related executions can remain mutually consistent even when the execution deviates from the intended problem or the obtained solution is not physically plausible.

%% file: body/motivating_example.tex
\section{Motivating Example}
\label{sec:motivating-example}

\begin{figure}[t]
  \centering
  \includegraphics[width=\linewidth]{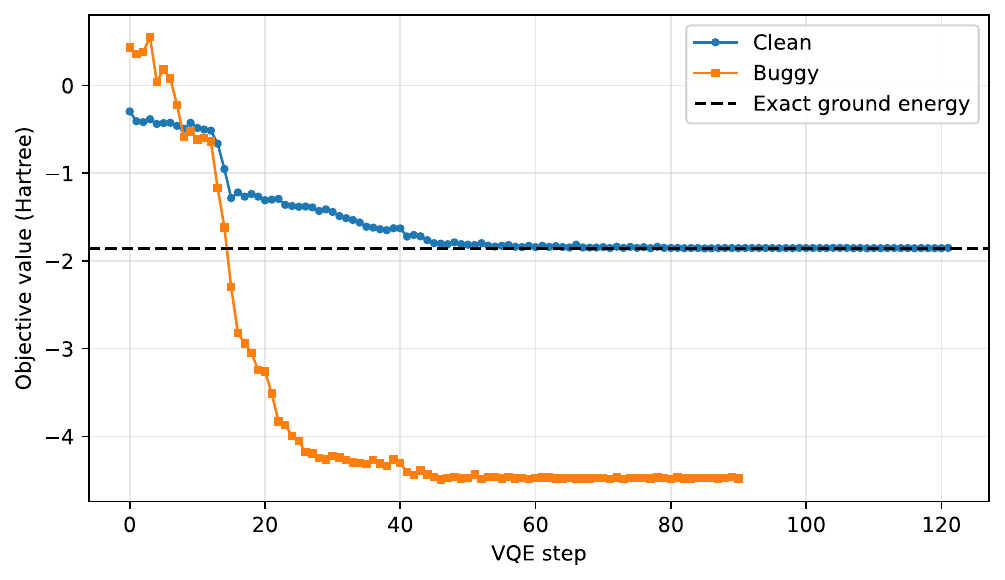}
  \caption{VQE convergence traces for the clean and buggy Hamiltonians in the~H$_2$ motivating example. The clean run optimizes the intended Hamiltonian~$H$, while the buggy run optimizes $H_{\mathrm{bug}} = H + 3.0\,YYII$.}
  \label{fig:motivating-example-convergence}
\end{figure}

As a motivating example, we consider a variational quantum eigensolver (VQE) workflow for estimating the ground-state energy of the H$_2$ molecule. The molecular geometry is H$(0,0,0)$--H$(0,0,0.735)$ with the STO-3G basis, and we consider an active-space reduction with two electrons in two spatial orbitals. The resulting electronic-structure problem is mapped to a four-qubit Hamiltonian by the Jordan-Wigner transformation. Let $H$ denote the mapped molecular Hamiltonian and let $\ket{\psi(\theta)}$ be the parameterized trial state prepared by the hardware-efficient ansatz shown in Fig.~\ref{fig:background-circuit}. The VQE task is to minimize the energy expectation value $\bra{\psi(\theta)} H \ket{\psi(\theta)}$ through repeated quantum measurements and classical parameter updates. The target Hamiltonian is:
\begin{align}
  H ={} & -0.811\,IIII
  + 0.172\,IIIZ
  - 0.226\,IIZI \nonumber \\
        & + 0.121\,IIZZ
  + 0.172\,IZII
  + 0.169\,IZIZ \nonumber \\
        & + 0.166\,IZZI
  + 0.045\,XXXX
  + 0.045\,XXYY \nonumber \\
        & + 0.045\,YYXX
  + 0.045\,YYYY
  - 0.226\,ZIII \nonumber \\
        & + 0.166\,ZIIZ
  + 0.175\,ZIZI
  + 0.121\,ZZII
  \label{eq:motivating-h2-hamiltonian}
\end{align}
Now suppose that, due to a user mistake or an implementation bug, an additional term is introduced into the Hamiltonian, and the VQE is executed with a buggy Hamiltonian $H_{\mathrm{bug}} = H + 3.0\,YYII$. As shown in Fig.~\ref{fig:motivating-example-convergence}, the optimization can still appear to converge normally. Nevertheless, the obtained solution is no longer physically plausible for the intended problem. This kind of failure is difficult to detect from the objective value alone, because apparent numerical convergence does not ensure that the obtained solution remains valid for the intended problem.

This example also clarifies why existing approaches are insufficient in this setting. If the correct solution were readily available, the bug could be detected by direct comparison, but such a reference cannot be obtained or verified classically in the regime where quantum advantage is expected. Assertion-based and property-based approaches would require detailed knowledge of the expected states, but the expected states during optimization are not known in advance, and the converged state is precisely what the algorithm is trying to find, so such knowledge cannot be specified in advance. Moreover, the execution infrastructure itself is functioning correctly, so approaches that compare related executions would not flag this apparently successful run. The core difficulty is that existing approaches do not verify whether the execution deviates from the intended problem or whether the obtained solution is physically plausible, which motivates a testing approach based on the problem structure itself.

%% file: body/proposed_method.tex
\section{Proposed Method}
\label{sec:proposed-method}

We describe the proposed method using the buggy-Hamiltonian case in Section~\ref{sec:motivating-example} as a running example.

\subsection{Overview}
\label{sec:proposed-method-overview}

The proposed method uses one or more operators $A$ satisfying $[H,A]=HA-AH=0$ as test specifications for the intended Hamiltonian $H$, where $H=\sum_i c_i P_i$ and $A=\sum_j d_j Q_j$ are expressed as linear combinations of Pauli strings with real coefficients $c_i$ and $d_j$. In quantum chemistry, particle number and total spin are natural choices because they are conserved in molecular systems: the corresponding operators commute with the molecular Hamiltonian, and their operator representations are known from the problem formulation. In the H$_2$ example of Section~\ref{sec:motivating-example}, we use the particle-number operator $N$ and the total-spin operator $S^2$, shown in Fig.~\ref{fig:h2-symmetry-operators}.

\begin{figure}[t]
  \centering
  \footnotesize
  \begin{tabular}{|c|}
    \hline
    \parbox{0.95\linewidth}{%
      \centering
      \setlength{\abovedisplayskip}{3pt}
      \setlength{\belowdisplayskip}{3pt}
      \setlength{\abovedisplayshortskip}{3pt}
      \setlength{\belowdisplayshortskip}{3pt}
      \[
        \begin{aligned}
          N ={} & 2\,IIII - 0.5\,IIIZ - 0.5\,IIZI \\
                & - 0.5\,IZII - 0.5\,ZIII
        \end{aligned}
      \]
      \[
        \begin{aligned}
          S^2 ={} & 0.75\,IIII - 0.375\,IZIZ - 0.125\,XYXY    \\
                  & - 0.125\,XXYY + 0.125\,YXXY - 0.125\,YYYY \\
                  & - 0.125\,XXXX + 0.125\,XYYX - 0.125\,YYXX \\
                  & - 0.125\,YXYX - 0.375\,ZIZI - 0.125\,IZZI \\
                  & + 0.125\,IIZZ + 0.125\,ZZII - 0.125\,ZIIZ
        \end{aligned}
      \]
    } \\
    \hline
  \end{tabular}
  \caption{Four-qubit Pauli decompositions of the particle-number operator $N$ and the total-spin operator $S^2$ used in the H$_2$ example.}
  \label{fig:h2-symmetry-operators}
\end{figure}

Using such operators, the proposed method validates a VQA through two complementary checks. The first is a health check~(HC), which monitors whether each execution step during optimization remains consistent with the symmetry of the intended Hamiltonian. The second is an acceptance check~(AC), which assesses the converged solution by examining whether the state is an eigenstate of the symmetry operators. Each check provides a necessary condition for consistency with the intended problem.

Figure~\ref{fig:sot-vqa} summarizes the overall workflow. At each optimization step, the health check is evaluated on the current state and the result is recorded (lines 5--6). After the optimizer converges, the acceptance check is evaluated on the converged state (line 11). When multiple symmetry operators are available, both checks are applied independently to each operator. We explain the two checks in turn below.

\begin{figure}[t]
  \centering
  \begin{algorithmic}[1]
    \REQUIRE Hamiltonian $H$; operator $A$ with $[H,A]=0$; parameterized circuit $U(\theta)$; initial parameters $\theta_0$; optimizer~$\mathcal{O}$; thresholds $\tau_{\mathrm{HC}},\tau_{\mathrm{AC}}$
    \ENSURE converged parameters $\theta^\ast$; converged objective value~$E^\ast$; health-check record $\mathcal{R}_{\mathrm{HC}}$; acceptance result $\mathrm{AC}_{A}(\psi^\ast) < \tau_{\mathrm{AC}}$
    \STATE Initialize $\theta \leftarrow \theta_0$ and $\mathcal{R}_{\mathrm{HC}} \leftarrow \emptyset$
    \REPEAT
    \STATE Prepare $\ket{\psi(\theta)} = U(\theta)\ket{\psi_0}$
    \STATE Estimate the objective value $\bra{\psi(\theta)} H \ket{\psi(\theta)}$
    \STATE Evaluate $\mathrm{HC}_{A}(\theta)$
    \STATE $\mathcal{R}_{\mathrm{HC}} \leftarrow \mathcal{R}_{\mathrm{HC}} \cup \{\mathrm{HC}_{A}(\theta) < \tau_{\mathrm{HC}}\}$
    \STATE Update $\theta$ using $\mathcal{O}$
    \UNTIL{the optimizer converges}
    \STATE $\theta^\ast \leftarrow \theta$
    \STATE Let $\ket{\psi^\ast} = U(\theta^\ast)\ket{\psi_0}$
    \STATE Evaluate $\mathrm{AC}_{A}(\psi^\ast)$ and $E^\ast = \bra{\psi^\ast}H\ket{\psi^\ast}$
  \end{algorithmic}
  \caption{Overall workflow of the proposed symmetry-oriented testing method for a VQA. When multiple symmetry operators are available, the health and acceptance checks are applied independently to each operator.}
  \label{fig:sot-vqa}
\end{figure}

\subsection{Health Check During Optimization}
\label{sec:proposed-method-health-check}

Let $U(\theta)$ be the parameterized quantum circuit used in the VQA, and let
\begin{equation}
  \ket{\psi(\theta)} = U(\theta)\ket{\psi_0}
\end{equation}
where $\ket{\psi_0}$ is the initial state. If the intended Hamiltonian $H$ commutes with the chosen operator $A$, then for any state $\ket{\phi}$,
\begin{equation}
  \bra{\phi} [H,A] \ket{\phi} = 0
\end{equation}
When the executed workflow deviates from the intended problem, the estimated value can depart from zero. We therefore define the health check at parameter $\theta$ by
\begin{equation}
  \mathrm{HC}_A(\theta)
  = \left| \bra{\psi(\theta)} HA \ket{\psi(\theta)}
  - \bra{\psi(\theta)} AH \ket{\psi(\theta)} \right|
\end{equation}
where the two terms are estimated separately through their Pauli decompositions:
\begin{align}
  \bra{\psi(\theta)} HA \ket{\psi(\theta)}
   & = \sum_{i,j} c_i d_j \bra{\psi(\theta)} P_iQ_j \ket{\psi(\theta)}
  \\
  \bra{\psi(\theta)} AH \ket{\psi(\theta)}
   & = \sum_{i,j} c_i d_j \bra{\psi(\theta)} Q_jP_i \ket{\psi(\theta)}
\end{align}
In an ideal execution, $\mathrm{HC}_A(\theta)$ vanishes for every parameter~$\theta$ visited during optimization. In practice, however, finite-shot estimation and hardware noise can make this quantity nonzero, so the optimization is judged acceptable at parameter $\theta$ when $\mathrm{HC}_A(\theta) < \tau_{\mathrm{HC}}$ for a small threshold $\tau_{\mathrm{HC}}$. Here, $\tau_{\mathrm{HC}}$ is chosen either as a statistically reasonable tolerance or by the user.

\begin{figure}[t]
  \centering
  \includegraphics[width=\linewidth]{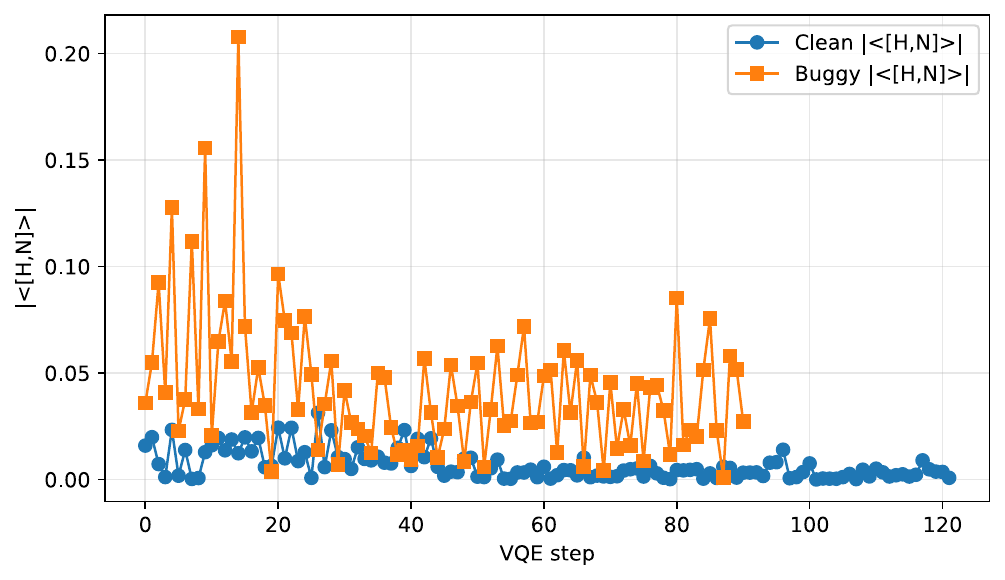}
  \caption{$\mathrm{HC}_N$ values for the clean and buggy runs in the H$_2$ motivating example across optimization iterations.}
  \label{fig:rq1-main}
\end{figure}

The H$_2$ motivating example illustrates this role. Since the intended Hamiltonian $H$ commutes with both $N$ and $S^2$, the check can be applied to both operators. Figure~\ref{fig:rq1-main} shows that, under the buggy Hamiltonian $H_{\mathrm{bug}}$, the $\mathrm{HC}_N$ trace clearly departs from the clean-run behavior, reaching values above 0.2. Since this example is evaluated on a simulator without hardware noise, this departure is too large to be plausibly explained by sampling noise alone. This indicates that the health check provides a practical basis for judging that the run is buggy, even when the objective alone appears well behaved.

\subsection{Acceptance Check for the Converged Solution}
\label{sec:proposed-method-acceptance-check}

The health check does not by itself guarantee that the converged solution is acceptable, so we also examine it directly through the \emph{variance} of the symmetry operator.

We apply this check when the converged state is expected to also be an eigenstate of $A$. One sufficient condition is that the target eigenvalue of $H$ be nondegenerate. Let $\ket{\psi_E}$ be an eigenstate of $H$ with a nondegenerate eigenvalue $E$, so that
\begin{equation}
  H\ket{\psi_E} = E\ket{\psi_E}
\end{equation}
Because $A$ commutes with $H$, we have
\begin{equation}
  H(A\ket{\psi_E}) = AH\ket{\psi_E} = E(A\ket{\psi_E})
\end{equation}
Hence $A\ket{\psi_E}$ is also an eigenvector of $H$ with the same eigenvalue. By nondegeneracy, it must be proportional to $\ket{\psi_E}$:
\begin{equation}
  A\ket{\psi_E} = a\ket{\psi_E}
\end{equation}
for some real scalar $a$, so the target state is also an eigenstate of $A$.

In quantum chemistry, nondegeneracy of the target eigenvalue is often known in advance from theory. For example, the ground state of a closed-shell molecule is commonly known to be nondegenerate.

If $\ket{\psi}$ is an eigenstate of $A$, repeated measurements of $A$ always return the same value, so the variance is zero. Conversely, a nonzero variance indicates that the state is not an eigenstate of $A$. We therefore use the variance of $A$ as the acceptance quantity for the converged state, defined as
\begin{equation}
  \mathrm{AC}_A(\psi) = \left|\mathrm{Var}_\psi(A)\right|
  = \left|\bra{\psi}A^2\ket{\psi} - \bra{\psi}A\ket{\psi}^2\right|
\end{equation}
Using the Pauli decomposition of $A$, the two terms can be expanded as
\begin{align}
  \bra{\psi}A^2\ket{\psi}
   & =
  \sum_{j,k} d_j d_k \bra{\psi}Q_jQ_k\ket{\psi}
  \\
  \bra{\psi}A\ket{\psi}^2
   & =
  \left(
  \sum_j d_j \bra{\psi}Q_j\ket{\psi}
  \right)^2
\end{align}

In an ideal case, $\mathrm{AC}_A(\psi)$ vanishes when the converged state is an eigenstate of $A$. In practice, finite-shot estimation and hardware noise can make this quantity nonzero, so we accept the converged solution when $\mathrm{AC}_A(\psi) < \tau_{\mathrm{AC}}$ for a small threshold $\tau_{\mathrm{AC}}$. Here, $\tau_{\mathrm{AC}}$ is chosen either as a statistically reasonable tolerance or by the user. When hardware noise is absent, a practical way to set $\tau_{\mathrm{AC}}$ is from the standard error:
\begin{equation}
  \sigma_{\mathrm{AC}}=\sqrt{\sigma_{\langle A^2\rangle}^2 + (2|\langle A\rangle|\sigma_{\langle A\rangle})^2}
\end{equation}
where $\sigma_{\langle A\rangle}$ and $\sigma_{\langle A^2\rangle}$ are the standard errors due to sampling noise in the estimates of $\langle A\rangle$ and $\langle A^2\rangle$, respectively. A simple practical choice is $\tau_{\mathrm{AC}} = 3\sigma_{\mathrm{AC}}$, which covers about 99.7\% of the sampling fluctuations under the usual Gaussian approximation.

Returning to the motivating example, because H$_2$ is a closed-shell molecule, its ground-state energy is nondegenerate. The target state is therefore expected to be an eigenstate of both $N$ and $S^2$, so the acceptance check is applicable to both operators. On a simulator without hardware noise and with $10{,}000$ shots, the clean run remains within the acceptable range, with $z=\mathrm{AC}_A(\psi)/\sigma_{\mathrm{AC}}=1.1$ for $N$ and $z=1.4$ for $S^2$. By contrast, the buggy run gives $z=50$ for $S^2$, far beyond the $3\sigma$ threshold ($z=3$). Thus, even though the buggy run appears to converge, the acceptance check reveals that the converged solution is no longer physically plausible for the intended Hamiltonian.

%% file: body/evaluation.tex
\section{Evaluation}
\label{sec:evaluation}

We evaluate the proposed framework through four experiment groups corresponding to the research questions in Section~\ref{sec:introduction}. Across these experiments, we apply the two diagnostics introduced in Section~\ref{sec:proposed-method}, the health check and the acceptance check, to VQA workflows. When the exact ground-state energy is reported, it is obtained by exact diagonalization of the target Hamiltonian and is used only for \textit{post hoc} analysis, not as a test oracle in the proposed framework. Energy values and energy differences are reported in Hartree. All experiments are conducted on a simulator using Qiskit 2.3.0 and qiskit-aer 0.17.2~\cite{qiskit}. Experimental code development supporting this section used Anthropic's Claude~\cite{anthropic_claude} and OpenAI's Codex~\cite{openai_codex}.

\subsection{RQ1: Can the Proposed Framework Provide Useful Fault-Detection Signals for Concrete Fault Scenarios in VQA Workflows Without Relying on the Exact Solution as a Test Oracle?}
\label{sec:evaluation-rq1}

This experiment applies the proposed framework to the H$_2$ instance in Section~\ref{sec:motivating-example} and several other injected faults without access to the correct answer as a test oracle. The results also help clarify how the proposed framework differs from relation-based testing~\cite{Wang2022QDiff,Paltenghi2023MorphQ,Linsey2024morphqplusplus}, which also do not require the correct answer. We do not include assertion-based or property-based testing, because using them to assess solution quality or to verify whether each execution step during optimization remains consistent with the intended problem would require detailed knowledge of the expected states at the relevant program points.

\subsubsection{Setup}
\label{sec:evaluation-rq1-setup}

The RQ1 experiments use the same H$_2$ instance as in Section~\ref{sec:motivating-example}, with the four-qubit Hamiltonian $H$ in Eq.~\eqref{eq:motivating-h2-hamiltonian} and the operators $N$ and $S^2$ shown in Fig.~\ref{fig:h2-symmetry-operators}.

For both the clean run and the buggy runs, we use the same hardware-efficient ansatz as in Fig.~\ref{fig:background-circuit}. Following standard VQA practice, each configuration is optimized from 50 initial points drawn uniformly from $[-0.8,0.8]$ with COBYLA for at most 200 iterations, and we report only the run with the lowest final energy. Expectation values are estimated by Pauli-term decomposition with 10,000 shots. To better approximate practical execution, the variational circuit and the measurement circuits are transpiled to the native gate set $\{\gate{CX}, \gate{RZ}, \gate{SX}\}$.

For the health check, we report the maximum $\mathrm{HC}_A(\theta)$ across all execution steps during optimization; for the acceptance check, we evaluate $\mathrm{AC}_A(\psi)$ at convergence using the $3\sigma$ threshold from Section~\ref{sec:proposed-method-acceptance-check}.

Under this common setup, the only difference lies in the injected fault. Table~\ref{tab:rq1-faults} summarizes the six faults used in RQ1, which span a broad range of levels, together with possible sources of each fault.

\begin{table}[t]
  \caption{Injected faults used in RQ1.}
  \label{tab:rq1-faults}
  \centering
  \scriptsize
  \begin{tabular}{p{0.23\columnwidth}p{0.135\columnwidth}p{0.493\columnwidth}}
    \hline
    Fault                  & Possible Source(s)    & Description                                                                                                                                                    \\
    \hline
    Stuck-zero readout     & Hardware / Middleware & All measurement outcomes are fixed to the all-zero bitstring.                                                                                                  \\
    Shifted wiring         & Compiler / Middleware & Each non-identity Pauli factor is cyclically shifted to the next qubit before basis-change insertion and measurement.                                          \\
    Randomized basis       & Compiler / Middleware & When constructing the measurement circuit for a Pauli term, the basis for each non-identity factor is chosen randomly from $\{X, Y, Z\}$.                      \\
    Wrong term weights     & Compiler / Middleware & In the Pauli decomposition $H = \sum_i c_i P_i$, all coefficients $c_i$ are replaced with~1.                                                                   \\
    Wrong Hamiltonian      & Compiler / User       & Optimization is performed on $H + 3.0\,\mathrm{YYII}$ instead of the intended Hamiltonian $H$ (same fault as in Section~\ref{sec:motivating-example}).         \\
    Underexpressive ansatz & User                  & The ansatz is restricted to two parameterized $\gate{R}_y$ gates on $q_0$ and $q_1$, followed by $\gate{CZ}$ gates from $q_0$ to each of the remaining qubits. \\
    \hline
  \end{tabular}
\end{table}

\input{body/generated/rq1_summary_table.tex}

\subsubsection{Results and Answer to RQ1}
\label{sec:evaluation-rq1-results}

Table~\ref{tab:rq1-summary} summarizes the results. The clean run provides the reference point: its exact-energy gap meets the criterion for chemical accuracy,\footnote{Chemical accuracy refers to an absolute energy error of approximately $1\,\mathrm{kcal/mol}$, or about $1.6\times 10^{-3}$ Hartree. It is a standard quality criterion in computational chemistry: energy differences computed within this tolerance are generally considered reliable for chemical predictions.} its health-check values during optimization are small, and its acceptance-check values at convergence remain within the $3\sigma$ threshold. This is the expected behavior.

Platform-level faults, including Stuck-zero readout, Shifted wiring, Randomized basis, and Wrong term weights, exhibit different combinations of health-check and acceptance-check responses. Randomized basis is the clearest case: $\max \mathrm{HC}_N$ rises from $3.13\times 10^{-2}$ in the clean run to $1.18$, and $\max \mathrm{HC}_{S^2}$ from $1.90\times 10^{-2}$ to $7.52\times 10^{-1}$, values far too large to be explained by finite-shot fluctuations, and the converged state also exceeds the $3\sigma$ acceptance threshold for both $N$ and $S^2$. Wrong term weights also shows health-check values exceeding $0.1$, too large to be attributed to finite-shot fluctuations, and both converged acceptance checks exceed the $3\sigma$ threshold. Shifted wiring shows milder health-check values during optimization but the converged $\mathrm{AC}_{S^2}$ still exceeds the $3\sigma$ threshold. Stuck-zero readout is distinctive in a different way. Both $\max \mathrm{HC}_N$ and $\max \mathrm{HC}_{S^2}$ collapse to $5.55\times 10^{-17}$ and $0$, values too close to zero to be plausible under finite-shot sampling. The converged $\mathrm{AC}$ values are also clearly abnormal: because every shot returns the all-zero bitstring, the estimated standard errors collapse to zero, yielding formally infinite $z$-scores. Taken together, these platform-level cases show that the two diagnostics respond differently across fault types, and are also natural targets for existing relation-based tests because they disturb the behavior of related executions.

Wrong Hamiltonian is especially important because it is not a platform-level fault. As shown in Table~\ref{tab:rq1-summary}, the obtained energy falls well below the exact ground-state energy; nonetheless, the convergence of the objective function alone does not reveal this fault (Fig.~\ref{fig:motivating-example-convergence}). The proposed framework, however, shows that the resulting run is clearly abnormal for the intended problem: $\max \mathrm{HC}_N$ exceeds $0.2$, too large to be attributed to finite-shot fluctuations, and the converged $\mathrm{AC}_{S^2}(\psi)$ exceeds the $3\sigma$ threshold decisively. Existing relation-based tests would not necessarily detect this mismatch, because these tests diagnose infrastructure correctness by comparing related executions, and the execution infrastructure itself functions correctly here.

Underexpressive ansatz highlights a related but distinct issue. As shown in Table~\ref{tab:rq1-summary}, the converged energy remains far above the exact ground-state energy since the ansatz is too restrictive to represent a plausible solution. Because this is not a fault in the execution flow itself, the health-check values during optimization remain close to zero. The acceptance check, however, reveals the problem: both $\mathrm{AC}_N(\psi)$ and $\mathrm{AC}_{S^2}(\psi)$ exceed the $3\sigma$ threshold at convergence, indicating that the converged solution is not physically plausible for the intended problem. Existing relation-based tests would also not expose this issue for the same reason: the execution infrastructure itself functions correctly.

\noindent\fbox{\begin{minipage}{\dimexpr\linewidth-2\fboxrule-2\fboxsep}
    \paragraph*{\textbf{Answer to RQ1}}
    Yes. The proposed framework provides useful fault-detection signals across a broad range of concrete fault scenarios without using the exact solution as a test oracle. Across the platform-level faults, problematic runs are exposed through the health check, the acceptance check, or both, depending on the fault type. Compared with relation-based tests, the clearest advantage appears in the last two cases, which are not platform-level faults. When optimization is carried out for the wrong Hamiltonian, related executions may remain mutually consistent, yet both checks still expose the mismatch. When the ansatz design is overly restrictive, the acceptance check still reveals the problem.
  \end{minipage}}

\subsection{RQ2: Does the Proposed Acceptance Check Show a Meaningful Correlation With Solution Quality Across Diverse Target Systems?}
\label{sec:evaluation-rq2}

This experiment evaluates whether the proposed acceptance check correlates with solution quality across diverse target systems. For each target system, we first search for a solution as close as possible to the optimum, then perturb it to generate many nearby states and examine how the acceptance-check values change as the energy error increases.

\subsubsection{Setup}
\label{sec:evaluation-rq2-setup}

The targets in this experiment are H$_2$, LiH, BeH$_2$, H$_2$O, and a 7-qubit Z2 lattice gauge theory model. H$_2$ uses the same 4-qubit active-space Hamiltonian as in RQ1. LiH is treated in the same way, with a 2-electron, 2-spatial-orbital active space that also maps to 4 qubits. BeH$_2$ uses a 4-electron, 3-spatial-orbital active space (6 qubits), and H$_2$O uses a 4-electron, 5-spatial-orbital active space in the 6-31G basis (10 qubits). In all four molecular cases, we use the same acceptance operators $N$ and $S^2$.

As the non-molecular benchmark, we use a 7-qubit Z2 lattice gauge model on two adjacent plaquettes with open boundaries. Unlike the molecular cases, where the relevant structure is given by global symmetries, this model is characterized by local gauge generators at the lattice vertices. Concretely, the six generators are
\[
  \begin{aligned}
    A_0 & = X_0 X_4, \quad
    A_1 = X_1 X_4, \quad
    A_2 = X_0 X_2 X_5,         \\
    A_3 & = X_1 X_3 X_5, \quad
    A_4 = X_2 X_6, \quad
    A_5 = X_3 X_6
  \end{aligned}
\]
With plaquette operators
\[
  B_0 = Z_0 Z_1 Z_4 Z_5,
  \qquad
  B_1 = Z_2 Z_3 Z_5 Z_6,
\]
the Hamiltonian used in this experiment is
\[
  H = -\left(B_0 + B_1 + A_0 + A_1 + A_2 + A_3 + A_4 + A_5\right)
\]

For every target system, we use the same hardware-efficient ansatz, consisting of three repeated blocks of $\gate{R}_y$ and $\gate{R}_z$ rotations on all qubits followed by $\gate{CX}$ gates between every pair of qubits, and a final layer of $\gate{R}_y$ and $\gate{R}_z$ rotations on all qubits. This yields 32, 32, 48, 80, and 56 variational parameters for H$_2$, LiH, BeH$_2$, H$_2$O, and Z2 LGT, respectively. This ansatz is slightly richer than the one used for H$_2$ in RQ1, reflecting the larger systems considered here. Optimization is performed with COBYLA for up to 3,000 iterations. For each target, we first run 1,000 optimizations from independent initial parameters and select the run with the best final energy. We then restart the optimization 1,000 times from nearby parameter settings around that run to search for reference parameters with better final energy. From those reference parameters, we generate 1,000 perturbed parameters by moving in random directions in parameter space, with distances sampled uniformly from 0.01 to 1.0. Expectation values are estimated with 10,000 shots.

For each perturbed state, we record the absolute energy error $|\Delta E|$ in Hartree, that is, the absolute difference from the exact ground-state energy, together with the corresponding acceptance-check values. In the molecular cases, these are $\mathrm{AC}_N(\psi)$ and $\mathrm{AC}_{S^2}(\psi)$, and in the lattice gauge model they are $\mathrm{AC}_{A_0}(\psi),\ldots,\mathrm{AC}_{A_5}(\psi)$; in both settings, we also take their maximum.

\begin{figure*}[t]
  \centering
  \begin{minipage}{0.32\textwidth}
    \centering
    {\footnotesize\textbf{$|\Delta E|$ vs.\ $\mathrm{AC}_N$}}\\
    \includegraphics[width=\linewidth]{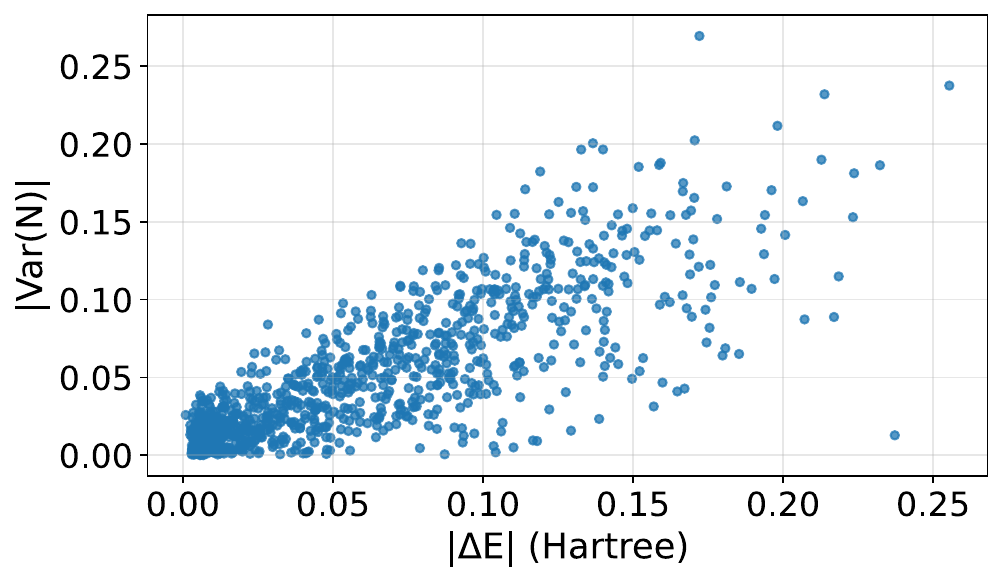}
  \end{minipage}\hfill
  \begin{minipage}{0.32\textwidth}
    \centering
    {\footnotesize\textbf{$|\Delta E|$ vs.\ $\mathrm{AC}_{S^2}$}}\\
    \includegraphics[width=\linewidth]{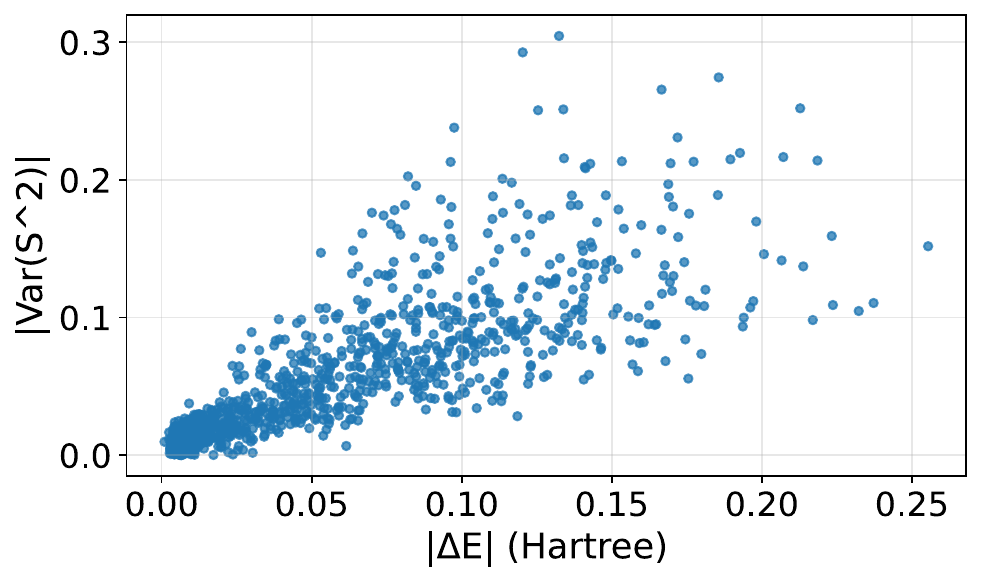}
  \end{minipage}\hfill
  \begin{minipage}{0.32\textwidth}
    \centering
    {\footnotesize\textbf{$|\Delta E|$ vs.\ $\max(\mathrm{AC}_N,\mathrm{AC}_{S^2})$}}\\
    \includegraphics[width=\linewidth]{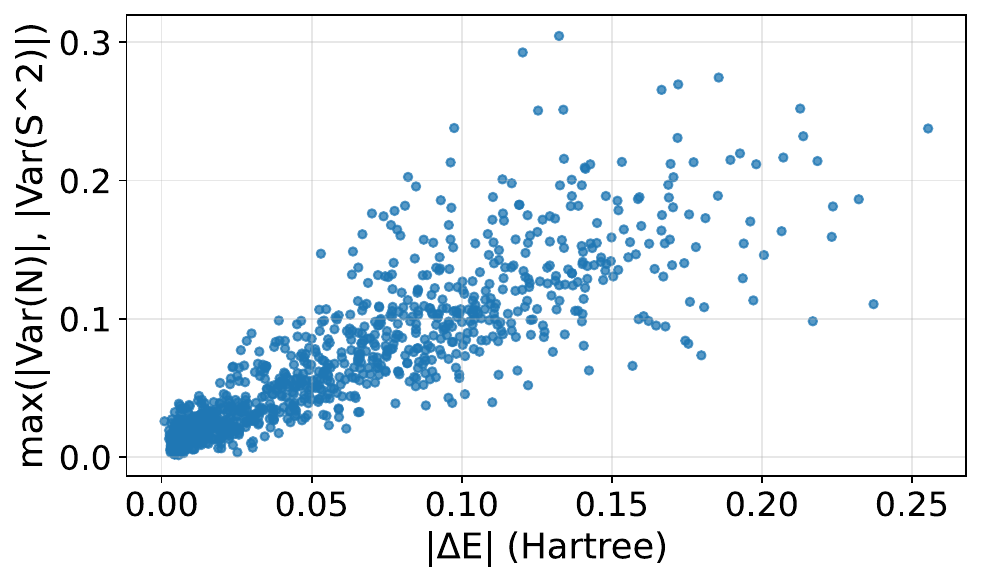}
  \end{minipage}
  \caption{Perturbation-scan results on H$_2$. Each point represents one perturbed parameter set near the reference solution.}
  \label{fig:rq2-scatter}
\end{figure*}

\subsubsection{Results and Answer to RQ2}
\label{sec:evaluation-rq2-results}

Table~\ref{tab:rq2-summary} shows the results. On H$_2$, both $\mathrm{AC}_N$ and $\mathrm{AC}_{S^2}$ are positively correlated with $|\Delta E|$, with Pearson correlations of 0.811 and 0.797, respectively. When they are combined as $\max(\mathrm{AC}_N,\mathrm{AC}_{S^2})$, the correlation increases to 0.857. Fig.~\ref{fig:rq2-scatter} visualizes this H$_2$ trend.

The other molecular systems support the same overall conclusion. Both operators carry useful information about solution quality: in each system, at least one achieves a Pearson correlation above 0.78, and which one is stronger varies. Combining them is advantageous; in particular, $\max(\mathrm{AC}_N,\mathrm{AC}_{S^2})$ gives the strongest correlation in H$_2$, BeH$_2$, and H$_2$O while remaining strongly correlated in LiH.

The Z2 lattice gauge model supports the same conclusion in a different physical setting. There, $\max(\mathrm{AC}_{A_0},\ldots,\mathrm{AC}_{A_5})$ has a correlation of 0.982 with $|\Delta E|$. The relationship between the acceptance check and solution quality is therefore not limited to molecular Hamiltonians or to the particular symmetries $N$ and $S^2$.

\input{body/generated/rq2_summary_table.tex}

\noindent\fbox{\begin{minipage}{\dimexpr\linewidth-2\fboxrule-2\fboxsep}
    \paragraph*{\textbf{Answer to RQ2}}
    Yes. The proposed acceptance check shows a meaningful correlation with solution quality across diverse target systems. On H$_2$, both $\mathrm{AC}_N$ and $\mathrm{AC}_{S^2}$ are positively correlated with $|\Delta E|$, and their maximum gives the strongest correlation. The same broad tendency persists across LiH, BeH$_2$, H$_2$O, and a Z2 lattice gauge model. These results show that the acceptance check provides informative signals about solution quality beyond a single benchmark or
    symmetry choice.
  \end{minipage}}

\subsection{RQ3: Can the Proposed Acceptance Check Serve as a Practical Filter for Low-Quality VQA Outcomes?}
\label{sec:evaluation-rq3}

RQ2 established that the proposed acceptance check is meaningfully correlated with solution quality. This experiment examines whether that relationship can be turned into a practical filter for low-quality outcomes in ordinary VQA runs without using the exact solution during execution. As a representative case, we study whether the converged acceptance-check values can identify poorer local optima.

\subsubsection{Setup}
\label{sec:evaluation-rq3-setup}

We use the same H$_2$ instance and the same acceptance operators $N$ and $S^2$ as in RQ1. We run the VQA 1,000 times from independent initial parameters drawn uniformly from $[-0.35,0.35]$, producing a pool of converged runs with varying quality. To make poorer local optima more likely, we intentionally use a simpler hardware-efficient ansatz. Concretely, it is the ansatz in Fig.~\ref{fig:background-circuit} with one $\gate{R}_y$ layer and one $\gate{CX}$ layer removed. On this 4-qubit problem, this gives 8 variational parameters, and each run is optimized with COBYLA for at most 300 iterations. Expectation values are estimated with 100,000 shots because this experiment evaluates filtering at the chemical-accuracy scale.

For each converged run, we record the absolute energy error $|\Delta E|$ together with $\max(\mathrm{AC}_N,\mathrm{AC}_{S^2})$. We then evaluate a two-threshold rule: $\tau_E$ defines which runs count as low quality, while $\tau_{\mathrm{AC}}$ is applied to $\max(\mathrm{AC}_N,\mathrm{AC}_{S^2})$ to decide which runs the filter rejects. Over a grid of $\tau_E$ and $\tau_{\mathrm{AC}}$ values, we measure recall, that is, the fraction of low-quality runs that are correctly rejected, and false-positive rate, that is, the fraction of runs satisfying the energy criterion that are incorrectly rejected. Here $\tau_E$ ranges from $1.60\times 10^{-3}$ to $1.84\times 10^{-2}$ Hartree and $\tau_{\mathrm{AC}}$ ranges from $6.0\times 10^{-3}$ to $1.0\times 10^{-2}$.

\subsubsection{Results and Answer to RQ3}
\label{sec:evaluation-rq3-results}

\begin{figure}[t]
  \centering
  \includegraphics[width=0.6\columnwidth]{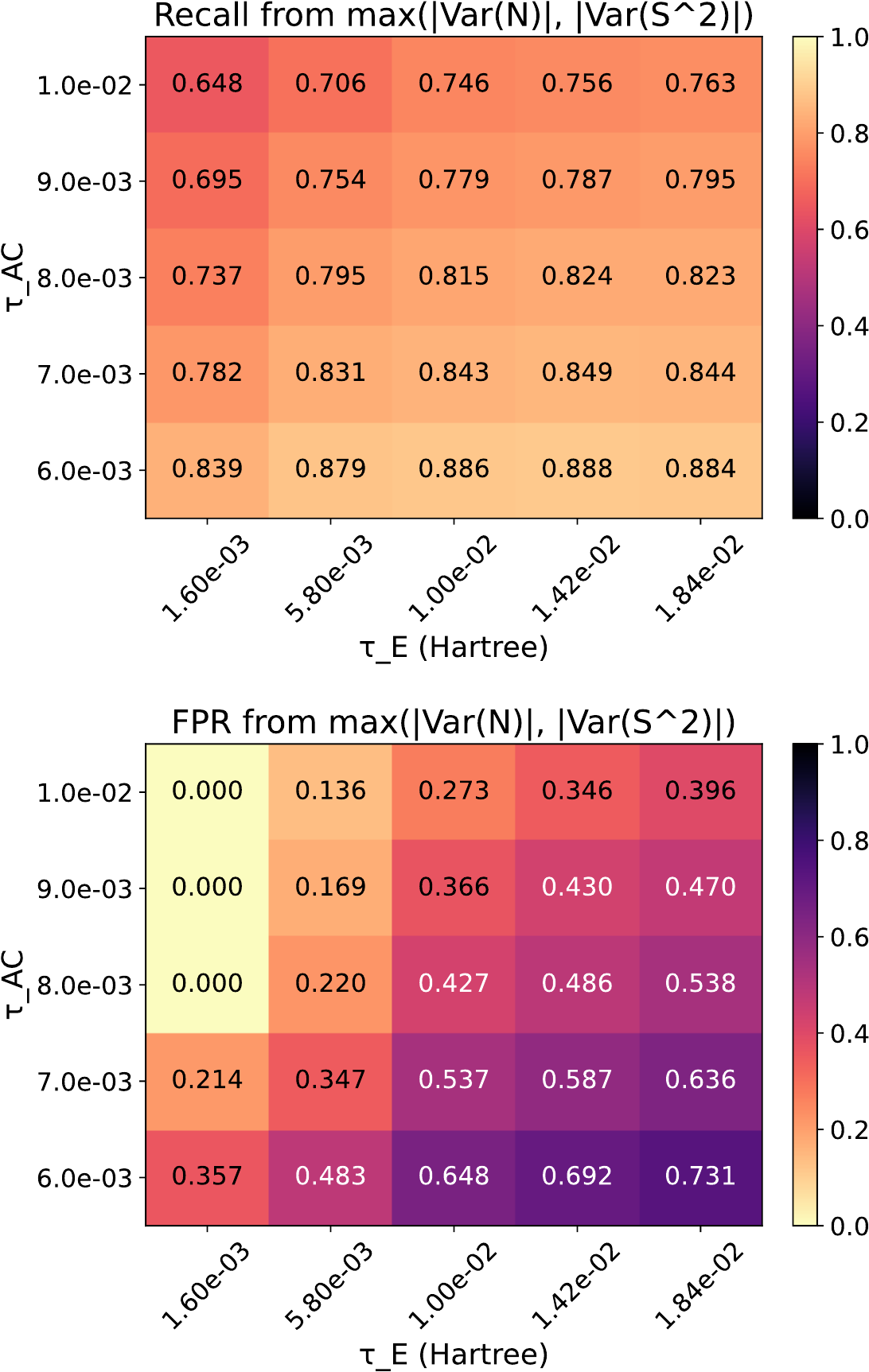}
  \caption{RQ3 screening heatmaps on 1,000 converged H$_2$ runs at 100,000 shots. The upper panel shows recall, and the lower panel shows the false-positive rate.}
  \label{fig:rq3-screening-heatmap}
\end{figure}

Fig.~\ref{fig:rq3-screening-heatmap} shows the results. The upper and lower panels report recall and false-positive rate, respectively, as functions of $\tau_E$ and $\tau_{\mathrm{AC}}$. The row corresponding to $\tau_E=1.60\times 10^{-3}$ Hartree is the chemical-accuracy setting of main interest here. Setting $\tau_{\mathrm{AC}}=8.0\times 10^{-3}$ successfully rejects 73.7\% of these runs while rejecting none of the runs that do satisfy the criterion.
Lowering $\tau_{\mathrm{AC}}$ to $7.0\times 10^{-3}$ makes the filter more aggressive, so recall rises to 78.2\%, but a 21.4\% false-positive rate also appears. Raising it to $9.0\times 10^{-3}$ or $1.0\times 10^{-2}$ makes the filter more conservative, keeping the false-positive rate at zero but reducing recall to 69.5\% and 64.8\%, respectively.

The same pattern holds at other values of $\tau_E$. As $\tau_E$ increases, fewer runs are classified as low quality, and the acceptance threshold that gives a favorable balance between recall and false-positive rate shifts to larger values of $\tau_{\mathrm{AC}}$.

\noindent\fbox{\begin{minipage}{\dimexpr\linewidth-2\fboxrule-2\fboxsep}
    \paragraph*{\textbf{Answer to RQ3}}
    Yes. The proposed acceptance check can serve as a practical filter for low-quality VQA outcomes. On the H$_2$ multi-start dataset at 100,000 shots, an appropriate acceptance threshold on $\max(\mathrm{AC}_N,\mathrm{AC}_{S^2})$ successfully rejects 73.7\% of the runs that do not satisfy chemical accuracy while producing no false positives. These results show that the correlation established in RQ2 can be turned into a practical rule for filtering lower-quality converged solutions, including local optima.
  \end{minipage}}

\subsection{RQ4: Can the Proposed Framework Provide Useful Diagnostic Signals Under Noisy Execution Conditions?}
\label{sec:evaluation-rq4}

This experiment evaluates whether the proposed framework remains informative under noisy execution conditions. Specifically, we increase the noise level and examine how the health check and the acceptance check change as the objective value degrades under noise.

\subsubsection{Setup}
\label{sec:evaluation-rq4-setup}

We use the same H$_2$ problem instance, ansatz, and symmetry operators as in RQ1. We first identify a noiseless reference run that achieves chemical accuracy, and then replay that run under different noise levels. For each replay, we reconstruct the objective value, the health-check values, and the acceptance-check values under the given noise level using the same sequence of variational parameters as in the reference run. Each Pauli term is estimated with 10,000 shots, and we keep the same transpilation policy as in RQ1. The inserted-noise model combines depolarizing gate noise on the $\gate{SX}$ and $\gate{CX}$ gates in the transpiled circuit with readout error; to reflect the usual implementation on superconducting devices, no inserted gate noise is applied to $\gate{RZ}$ gates. The noise level is scaled from 0$\times$ to 10$\times$. In the reference 1$\times$ profile, the single-qubit gate error rate is 0.0001 on $\gate{SX}$ gates, the two-qubit gate error rate is 0.001 on $\gate{CX}$ gates, and the readout error rates are $P(0\mid 1)=0.002$ and $P(1\mid 0)=0.001$. Note that the 0$\times$ case still differs from the noiseless baseline because it includes sampling noise from finite shots.

\begin{figure*}[t]
  \centering
  \begin{minipage}{0.32\textwidth}
    \centering
    {\footnotesize\textbf{Objective value}}\\
    \includegraphics[width=\linewidth]{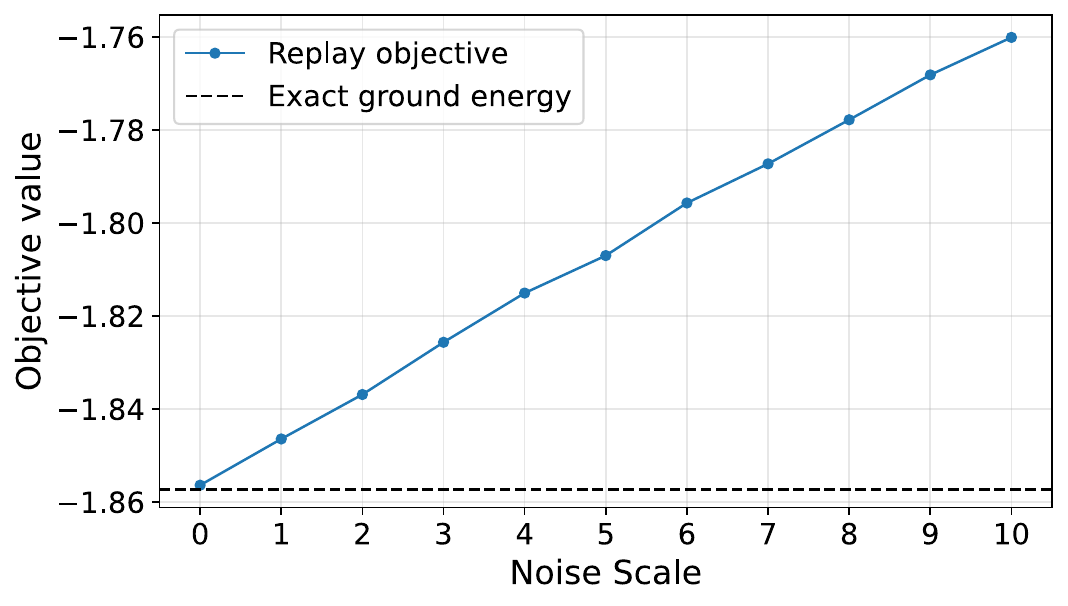}
  \end{minipage}\hfill
  \begin{minipage}{0.32\textwidth}
    \centering
    {\footnotesize\textbf{Health-check values}}\\
    \includegraphics[width=\linewidth]{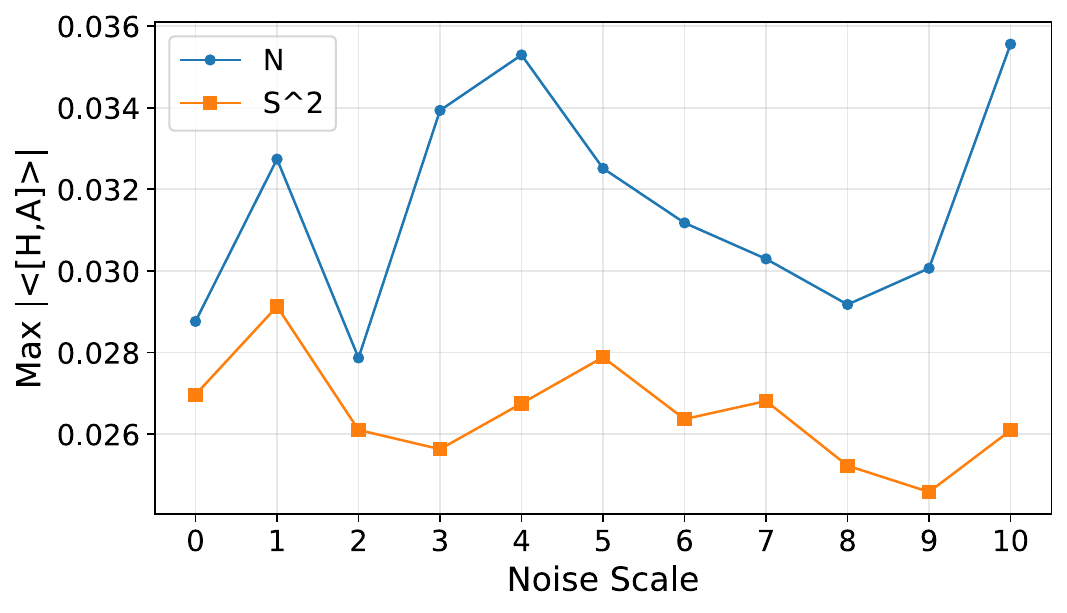}
  \end{minipage}\hfill
  \begin{minipage}{0.32\textwidth}
    \centering
    {\footnotesize\textbf{Acceptance-check values}}\\
    \includegraphics[width=\linewidth]{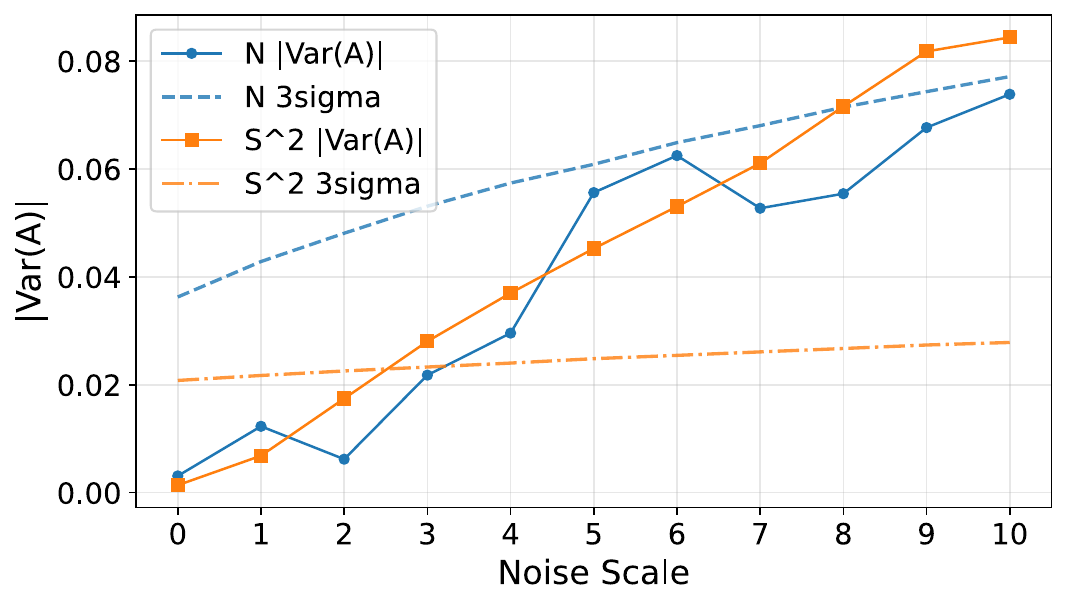}
  \end{minipage}
  \caption{RQ4 noisy-execution results on the H$_2$ VQA. The three panels show the objective value, the health-check values, and the acceptance-check values as the inserted-noise level is increased from 0$\times$ to 10$\times$. Here, the reference 1$\times$ profile uses single-qubit gate error 0.0001 on $\gate{SX}$ gates, two-qubit gate error 0.001 on $\gate{CX}$ gates, and readout error rates $P(0\mid 1)=0.002$ and $P(1\mid 0)=0.001$.}
  \label{fig:rq4-noise-scaling}
\end{figure*}

\subsubsection{Results and Answer to RQ4}
\label{sec:evaluation-rq4-results}

Fig.~\ref{fig:rq4-noise-scaling} shows the results of this experiment. The left panel shows that the objective value deteriorates steadily as the inserted-noise level is increased. The gap from the exact ground-state energy grows from $8.42\times 10^{-4}$ at 0$\times$ to $9.71\times 10^{-2}$ at 10$\times$, so the replayed run becomes progressively less accurate under stronger noise.

The acceptance check tracks this loss of accuracy clearly. As the energy error increases under noise, $\mathrm{AC}_{S^2}$ rises quickly and continues to grow across the higher-noise runs, with its $z$-score exceeding 3 from 3$\times$ noise onward. $\mathrm{AC}_N$ also steadily increases. These trends show that the acceptance check is sensitive to the noise-induced increase in energy error and can therefore serve as an indicator of noise-induced degradation.

The health check remains close to zero across the noise-scaling study, consistent with its commutator-based definition: the inserted noise degrades the objective value but does not alter the Hamiltonian or its symmetry relations, and therefore does not lead to large violations of $\langle [H,A]\rangle = 0$ in this setting.

\noindent\fbox{\begin{minipage}{\dimexpr\linewidth-2\fboxrule-2\fboxsep}
    \paragraph*{\textbf{Answer to RQ4}}
    \revised{Yes. The proposed framework still provides useful diagnostic signals under noisy execution. In the H$_2$ noise-scaling study, the energy error grows steadily as inserted noise is increased, and the acceptance check tracks this trend. This suggests that the acceptance check can serve as an indicator of noise-induced degradation of solution quality.} In addition, the health check remains close to zero, indicating that noise alone does not spuriously indicate a consistency violation when the noise does not alter the underlying symmetry relations.
  \end{minipage}}

%% file: body/generated/rq1_summary_table.tex
\begin{table*}[t]
  \caption{RQ1 results for the H$_2$ clean run and the injected faults in Table~\ref{tab:rq1-faults}.}
  \label{tab:rq1-summary}
  \centering
  \scriptsize
  \begin{tabular}{@{}lrrr r@{\,}l r@{\,}l@{}}
    \hline
    Case & \multicolumn{1}{c}{$\Delta E_{\mathrm{exact}}$} & \multicolumn{1}{c}{Max $\mathrm{HC}_N$} & \multicolumn{1}{c}{Max $\mathrm{HC}_{S^2}$} & \multicolumn{2}{c}{$\mathrm{AC}_N$ ($z$ value)} & \multicolumn{2}{c}{$\mathrm{AC}_{S^2}$ ($z$ value)} \\
    \hline
    Clean run & $-2.44\times 10^{-4}$ & $3.13\times 10^{-2\phantom{0}}$ & $1.90\times 10^{-2\phantom{0}}$ & $1.57\times 10^{-2}$ & $(1.1\times 10^{0\phantom{-}})$ & $1.01\times 10^{-2}$ & $(1.4\times 10^{0})$ \\
    Stuck-zero readout & $+2.04\times 10^{0\phantom{-}}$ & $5.55\times 10^{-17}$ & $0.00\times 10^{0\phantom{-0}}$ & $0.00\times 10^{0\phantom{-}}$ & $(\infty)$ & $1.25\times 10^{0\phantom{-}}$ & $(\infty)$ \\
    Shifted wiring & $+1.20\times 10^{-2}$ & $3.81\times 10^{-2\phantom{0}}$ & $2.61\times 10^{-2\phantom{0}}$ & $1.40\times 10^{-3}$ & $(1.4\times 10^{-1})$ & $2.59\times 10^{-2}$ & $(3.5\times 10^{0})$ \\
    Randomized basis & $+4.72\times 10^{-1}$ & $1.18\times 10^{0\phantom{-0}}$ & $7.52\times 10^{-1\phantom{0}}$ & $1.29\times 10^{-1}$ & $(3.1\times 10^{0\phantom{-}})$ & $1.84\times 10^{-1}$ & $(1.1\times 10^{1})$ \\
    Wrong term weights & $-3.16\times 10^{0\phantom{-}}$ & $1.27\times 10^{-1\phantom{0}}$ & $1.14\times 10^{-1\phantom{0}}$ & $2.10\times 10^{0\phantom{-}}$ & $(4.5\times 10^{1\phantom{-}})$ & $2.87\times 10^{1\phantom{-}}$ & $(3.9\times 10^{2})$ \\
    Wrong Hamiltonian & $-2.63\times 10^{0\phantom{-}}$ & $2.08\times 10^{-1\phantom{0}}$ & $8.19\times 10^{-2\phantom{0}}$ & $1.84\times 10^{-2}$ & $(4.4\times 10^{-1})$ & $7.40\times 10^{-1}$ & $(5.0\times 10^{1})$ \\
    Underexpressive ansatz & $+5.95\times 10^{-1}$ & $2.24\times 10^{-2\phantom{0}}$ & $2.03\times 10^{-2\phantom{0}}$ & $9.31\times 10^{-2}$ & $(7.1\times 10^{0\phantom{-}})$ & $9.62\times 10^{-2}$ & $(8.5\times 10^{0})$ \\
    \hline
  \end{tabular}
  \par\vspace{2pt}
  $\Delta E_{\mathrm{exact}}$: gap from the exact energy. An entry of $\infty$ indicates division by zero in the corresponding $z$-value calculation.\par
\end{table*}

%% file: body/generated/rq2_summary_table.tex
\begin{table}[t]
  \caption{RQ2 correlations between the acceptance check and solution quality across target systems.}
  \label{tab:rq2-summary}
  \centering
  \scriptsize
  \begin{tabular}{lccc}
    \hline
    System  & Corr.\ with $\mathrm{AC}_N$ & Corr.\ with $\mathrm{AC}_{S^2}$ & Corr.\ with max \\
    \hline
    H$_2$ & 0.811 & 0.797 & 0.857 \\
    LiH & 0.874 & 0.668 & 0.831 \\
    BeH$_2$ & 0.797 & 0.877 & 0.891 \\
    H$_2$O & 0.783 & 0.835 & 0.837 \\
    Z2 LGT & --- & --- & 0.982 \\
    \hline
  \end{tabular}
  \par\vspace{2pt}
  max: $\max(\mathrm{AC}_N,\mathrm{AC}_{S^2})$ for the molecular cases and $\max(\mathrm{AC}_{A_0},\ldots,\mathrm{AC}_{A_5})$ for the lattice gauge model.\par
\end{table}

%% file: body/discussion.tex
\section{Discussion}
\label{sec:discussion}

\subsection{Relation to Prior Uses of Symmetry}
\label{sec:discussion-relation-to-prior-uses-of-symmetry}

The proposed framework uses symmetry to construct test oracles for VQAs. Prior work has used symmetry in a different role. Symmetry-preserving ansatz~\cite{Gard2020SymmetrypreservingAnsatz} encodes symmetry directly into the circuit so that the generated states remain within the target symmetry sector, thereby reducing the search space. The trade-off is that enforcing more symmetries often requires additional entangling gates, which can increase circuit depth and noise sensitivity on noisy intermediate-scale quantum (NISQ) hardware. The two approaches are therefore complementary: some symmetries can be built into the ansatz when the circuit cost is acceptable, whereas others can be checked through testing. Embedding symmetry into the circuit is also not itself a means of diagnosing faults.

Symmetry-adapted VQE~\cite{Seki2020SymmetryAdoptedVQE} modifies the energy functional so that the optimizer minimizes the energy of the trial state projected onto the target symmetry sector rather than the energy of the full trial state; because the projection is not unitary, it is handled classically. This guides the optimization toward variational states with larger overlap with the desired symmetry subspace. Symmetry verification~\cite{Sagastizabal2019SymmetryErrorMitigation} takes a different approach: the VQE runs without modification, and after convergence the measurement results are classically postprocessed to project the converged state onto the target symmetry sector, removing symmetry-violating components and yielding an error-mitigated energy estimate.

These approaches use symmetry to recover or improve the outcome of a run, whereas the proposed framework uses symmetry to diagnose whether the run should be trusted.

\subsection{Scope and Limitations}
\label{sec:discussion-practical-scope-and-limitations}

\subsubsection{Diagnostic Meaning}
The proposed framework provides necessary rather than sufficient conditions for reliable VQA behavior: passing the health and acceptance checks does not guarantee that a VQA has reached the exact target state. \revised{For example, in the screening experiment in RQ3, some runs that do not satisfy chemical accuracy pass the acceptance threshold.}

\revised{In practice, however, the evaluation shows that both checks are effective.} In RQ1, the health check and the acceptance check together detect all six injected faults, which span diverse fault types from platform-level errors to user-side design issues. In RQ2, the acceptance check is positively correlated with energy error across several molecular systems and a Z2 lattice gauge model, and \revised{in RQ3 it rejects 73.7\% of low-quality converged runs on H$_2$ with no false positives.} In RQ4, the acceptance check tracks noise-induced loss of accuracy.

\subsubsection{Operator Choice and Applicability}
The framework depends on choosing symmetry operators that are appropriate for the target problem. In domains such as quantum chemistry, symmetry information is already part of the physical problem description and is routinely used in algorithm design (Section~\ref{sec:discussion-relation-to-prior-uses-of-symmetry}). \revised{In lattice gauge theories, local gauge generators commute with the Hamiltonian by construction (Section~\ref{sec:evaluation-rq2}). For other problem domains, identifying appropriate commuting operators may require domain-specific analysis.}

For the acceptance check, additional care is needed only in the degenerate case. As discussed in Section~\ref{sec:proposed-method-acceptance-check}, when the target eigenvalue is nondegenerate, any operator commuting with the Hamiltonian can be used, because the target state is then automatically an eigenstate of it. In the degenerate case, by contrast, one must identify commuting operators for which the target state is still expected to be an eigenstate. Identifying such operators across a wider range of applications is an important research direction.

\begin{revisedblock}
  \subsubsection{Validation of the Commutation Assumption}
  When the diagnostics report an anomaly, the framework cannot distinguish whether the cause lies in the test specification $A$ or in the executed workflow. This is a general limitation of specification-based testing. In practice, this can be mitigated by checking $[H, A] = 0$ algebraically before execution.
\end{revisedblock}

\subsubsection{Measurement Cost}
The main practical issue is the additional Pauli terms whose expectation values must be estimated for the health and acceptance checks. \revised{For the H$_2$ Hamiltonian used in this study, which consists of 15 Pauli terms, the health and acceptance checks for $N$ and $S^2$ require 9 additional terms in total.} This overhead can be mitigated in several standard ways. Because commuting Pauli operators can be estimated together, Pauli grouping reduces the measurement overhead associated with these additional terms~\cite{Verteletskyi_2020_PauliGroup}. Classical-shadow methods likewise provide a way to predict many Pauli observables, with a number of measurements that grows only logarithmically with the number of Pauli observables~\cite{Huang_2020}. In addition, the health check need not be applied at each execution step during optimization and can instead be monitored periodically.

\subsection{Threats to Validity}
\label{sec:discussion-threats-to-validity}

The main validity concern is whether the reported effects depend too strongly on the particular experimental choices.

For RQ1, the main threat is that the evaluation uses artificially injected faults and therefore does not cover all faults that arise in practice. However, the injected faults were chosen to span distinct levels, from platform-level errors such as wiring and readout faults to user-side issues such as Hamiltonian specification and ansatz design, and include cases where the execution infrastructure itself functions correctly. In addition, RQ3 shows that the acceptance check is useful beyond injected-fault scenarios, serving as a practical filter for low-quality converged runs.

For the acceptance check, the main concern is how far the correlation observed in RQ2 extends beyond the target systems studied here. RQ2 provides empirical support on this point by showing that the acceptance check is useful across multiple molecular systems with different target Hamiltonians and qubit counts. \revised{The same correlation is observed} in a Z2 lattice gauge model, indicating that its usefulness is not tied to a particular application domain or symmetry class. A further limitation is that the perturbation scan in RQ2 probes only the neighborhood of a single reference solution for each target system. RQ3 provides complementary evidence by showing that the acceptance check serves as a practical filter on independently converged runs at different local optima.

For noisy execution, the main threat is that RQ4 is not an ordinary noisy optimization experiment. Instead, it replays a fixed sequence of variational parameters obtained in the noiseless setting. However, this does not pose a major limitation for the proposed diagnostics themselves, because the health check verifies symmetry consistency at each execution step regardless of how the parameters were chosen, and the acceptance check depends only on the converged state. By holding \revised{the parameter sequence fixed, the experiment isolates the effect of noise from differences in the optimization path.} Another threat is that RQ4 does not evaluate the effect of error-mitigation techniques~\cite{Temme2017ErrorMitigation}, although the proposed diagnostics are compatible with such techniques because they are defined through expectation values.

%% file: body/conclusion.tex
\section{Conclusion}
\label{sec:conclusion}

In this paper, we addressed the oracle problem in testing variational quantum algorithms by proposing a symmetry-oriented testing framework. The framework uses operators that commute with the target Hamiltonian as test specifications and combines two diagnostics: a health check that verifies symmetry consistency at each execution step during optimization, and an acceptance check that examines whether the converged state is an eigenstate of the symmetry operators. This enables testing of VQA runs without requiring the exact solution as a test oracle.

The evaluation demonstrated that the two checks together detect diverse faults, from platform-level errors to user-side design issues. The acceptance check is positively correlated with solution quality across multiple molecular systems and a Z2 lattice gauge model, and can be turned into a practical filter for low-quality converged runs. Under noisy execution conditions, the acceptance check tracks noise-induced degradation of solution quality. This work takes a first step toward testing quantum software in regimes where quantum computers provide a computational advantage and exact classical solutions are unavailable as test oracles.

%% file: body/data_availability.tex
\section*{Data Availability}

\revised{The implementation and experimental artifacts are available} at \revised{\url{https://doi.org/10.5281/zenodo.21405151}} or \revised{\url{https://github.com/FujitsuResearch/SOT-Framework}}.

%% file: body/acknowledgment.tex
\section*{Acknowledgment}

The authors used Anthropic's Claude~\cite{anthropic_claude} and OpenAI's Codex~\cite{openai_codex} \revised{for drafting and language refinement in all sections, and for} experimental code development supporting Section~\ref{sec:evaluation}.
The authors reviewed all AI-generated content and take full responsibility for the final manuscript.